\documentstyle[11pt,newpasp,psfig]{article}
\begin{document}

\markboth{Charmandaris, et al.}{Mid-IR imaging of Toomre's Merger Sequence}
\setcounter{page}{1}

\title{Mid-IR imaging of Toomre's Merger Sequence}
\author{V. Charmandaris}
\affil{DEMIRM, Observatoire de Paris, 61 Av. de l'Observatoire,
F-75 014, Paris, France}
\author{O.~Laurent, I.F. Mirabel, P. Gallais}
\affil{CEA/DSM/DAPNIA, Service d'Astrophysique, F-91191 Gif-sur-Yvette, France}
\begin{abstract}

We present results on mid-IR (5--16$\mu m$) spectral imaging of a
sequence of interacting galaxies, observed by ISOCAM. The galaxies are
part of the well known Toomre's ``merger sequence'' which was defined
as a sample of galaxies depicting progressive snapshots in the time
evolution of a merging event. To trace the intensity of the radiation
field in a starburst, we use the ratio of the 15$\mu m$ to 7$\mu m$
flux. Our analysis indicates that this ratio increases as galaxies
move from the pre-starburst to the starburst phase and goes again down
to $\sim$ 1 in the post-starburst phase, a value typical of normal
star forming regions in galactic disks. Moreover, we find that the
variation of this ratio is well correlated with the one of the IRAS
25$\mu m$/12$\mu m$ and 60$\mu m$/100$\mu m$ flux ratios.
\end{abstract}


\vspace*{-0.5cm}
\section{Introduction}
\vspace*{-0.2cm}

One of the major steps in the understanding of galaxy evolution was
the realization that tails and bridges are the result of galaxy
interactions (Toomre \& Toomre 1972). The subsequent proposal by
Toomre (1977) of using the morphology of the tidal features to create
a ``merging sequence'' of 11 NGC galaxies triggered numerous
multi-wavelength studies of those systems (i.e. Hibbard 1995). As a
result, several observational characteristics have been tested as
alternatives of assigning an ``age'' to the event of the interaction
(i.e. Schweizer 1998). Moreover, the discovery by IRAS of the class of
luminous IR galaxies and the revelation later on that they are also
interacting/merging systems, attracted further attention to this
problem (see Sanders \& Mirabel 1996 for a review).

\vspace*{-0.2cm}
\section{Discussion: The Mid-IR perspective}
\vspace*{-0.2cm}

To improve our knowledge of the properties of interacting galaxies in
the mid-IR, we used ISOCAM to perform deep spectral imaging
observations in the (5--16$\mu m$) of sample including most of the
well known nearby active/interacting systems (Laurent et
al. 1999). The analysis of the spectral characteristics of our sample
revealed that in galaxies where an active nucleus does not have a
detectable contribution to their mid-IR flux, one can use the flux
ratio of LW3(12-18\,$\mu m$) to LW2(5-8.5\,$\mu m$) ISOCAM filters as
an indicator of the intensity of the star formation activity. This
ratio samples the mid-IR continuum emission originating from very
small dust grains (radius\,$<$\,10\,nm) heated to high temperatures
due to their close proximity to OB stars (D\'esert et al.  1990).

\vspace*{-0.2cm}
\begin{figure}[ht]
\centerline{\psfig{figure=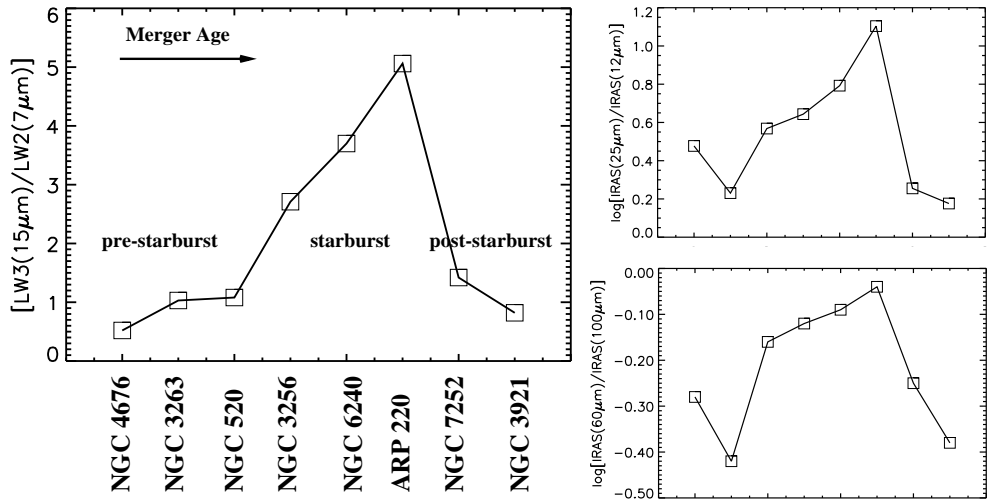,width=5.25in}}
\vspace*{-0.2cm}
\caption{A comparison of the variation of the ISOCAM LW3/LW2 flux ratio
along the merging sequence, with the well known IRAS flux ratios. Note
how well the ISOCAM starburst diagnostic traces the evolution of
the star forming activity/merger age of the sequence. The IRAS 12$\mu
m$ and 25$\mu m$ fluxes, corrected for the extent of the galaxies,
have been kindly provided by D.B. Sanders (Univ.  Hawaii).}
\label{fig1}
\end{figure}

In Fig 1. we present eight galaxies of our sample found in increasing
stages of interaction: from NGC\,4676, to Arp\,220, and NGC\,7252. We
observe that the ISOCAM LW3/LW2 diagnostic traces the star formation
activity in the galaxies and that it is well correlated with the
corresponding IRAS flux ratios (Charmandaris et al. 1999). This
suggests that even though the bolometric luminosity of luminous
infrared galaxies is found at $\lambda \geq 40 \mu m$, the study of the
mid-IR spectral energy distribution is a powerful tool in understanding
their global star formation history.


\vspace*{-0.5cm}
\begin{references}
\vspace*{-0.2cm}
{\small
\reference Charmandaris, V., Laurent, O., Mirabel, I.F.,  et al. 1999, (in preparation)
\reference D\'esert, F.-X, Boulanger, F., \& Puget, J.L. 1990,  \aap, 237, 215
\reference Laurent, O., Mirabel, I.F., Charmandaris, V., et al. 1999, \aap. (submitted)
\reference Hibbard, J., 1995 Ph.D. Thesis, Columbia Univ.
\reference Sanders, D.B., \& Mirabel, I.F. 1996, \araa, 34, 749
\reference Schweizer,  F. 1998, in 
	``Galaxies: Interactions and Induced Star Formation'', 
	Saas-Fee Advanced Course 26, 105  
\reference Toomre, A. 1977, in 
	``The Evolution of Galaxies and Stellar Population'', 
	eds. B.M. Tinsley \& R.B. Larson,  401
\reference Toomre, A., \& Toomre, J. 1972, \apj, 405, 142

}
\end{references}
\end{document}